%% file: main.tex
\documentclass[10pt,journal,compsoc]{IEEEtran}

\ifCLASSOPTIONcompsoc
  \usepackage[nocompress]{cite}
\else
  \usepackage{cite}
\fi

\ifCLASSINFOpdf
  \usepackage[pdftex]{graphicx}
  \graphicspath{{./figures/}}
  \DeclareGraphicsExtensions{.pdf,.jpeg,.png}
\else
\fi

\usepackage{amssymb,amsmath,amsfonts}
\usepackage{multirow}

\usepackage{algorithm}
\usepackage{algorithmic}

\begin{document}
%
\title{Collective Semi-Supervised Learning for User Profiling in Social Media}
%
%
%
%

\author{Richard J. Oentaryo,~\IEEEmembership{}
        Ee-Peng Lim,~\IEEEmembership{}
        Freddy Chong Tat Chua,~\IEEEmembership{}
        Jia-Wei Low,~\IEEEmembership{}
        and David Lo~\IEEEmembership{}
\IEEEcompsocitemizethanks{\IEEEcompsocthanksitem R. J. Oentaryo, E.-P. Lim, and D. Lo are with the Living Analytics Research Centre, Singapore Management University, Singapore 178902. E-mail: \{roentaryo, eplim, davidlo\}@smu.edu.sg\protect\\
\IEEEcompsocthanksitem F. C. T. Chua is with the Mechanism and Design Lab, HPLabs, Palo Alto, CA 94304, USA. E-mail: freddy.chua@hp.com\protect\\
\IEEEcompsocthanksitem J.-W. Low is with the Infocomm Development Authority, 10 Pasir Panjang Road, Singapore 117438. E-mail: davidlowjw@gmail.com 
}
\thanks{Manuscript received Month XX, 2015; revised Month XX, 2015.}}

\IEEEtitleabstractindextext{%
\begin{abstract}
The abundance of user-generated data in social media has incentivized the development of methods to infer the latent attributes of users, which are crucially useful for personalization, advertising and recommendation. However, the current user profiling approaches have limited success, due to the lack of a principled way to integrate different types of social relationships of a user, and the reliance on  scarcely-available labeled data in building a prediction model. In this paper, we present a novel solution termed \emph{Collective Semi-Supervised Learning} (CSL), which provides a principled means to integrate different types of social relationship and unlabeled data under a unified computational framework. The joint learning from multiple relationships and unlabeled data yields a computationally sound and accurate approach to model user attributes in social media. Extensive experiments using Twitter data have demonstrated the efficacy of our CSL approach in inferring user attributes such as account type and marital status. We also show how CSL can be used to determine important user features, and to make inference on a larger user population.
\end{abstract}

}

\maketitle

\IEEEdisplaynontitleabstractindextext

%
\IEEEpeerreviewmaketitle

\input{introduction}
\input{related_work}
\input{framework}
\input{problem}
\input{experiment}
\input{conclusion}



\ifCLASSOPTIONcompsoc
  \section*{Acknowledgments}
\else
  \section*{Acknowledgment}
\fi

This work is supported by the Singapore National Research Foundation under its International Research Centre @ Singapore Funding Initiative and administered by the IDM Programme Office, Media Development Authority (MDA).

\ifCLASSOPTIONcaptionsoff
  \newpage
\fi

\bibliographystyle{IEEEtran}
\bibliography{IEEEabrv,references}

%



\begin{IEEEbiography}{Richard J. Oentaryo}
is a Research Scientist at the Living Analytics Research Centre, School of Information Systems, Singapore Management University. He obtained his Ph.D. and B.Eng. (First Class Honour) in Computer Engineering from the Nanyang Technological University, Singapore, in 2011 and 2004 respectively. His research interests include machine learning, data mining, and nature-inspired computing. Dr. Oentaryo has published more than 25 papers in various international journals and conferences. 
\end{IEEEbiography}

\begin{IEEEbiography}{Ee-Peng Lim}
is a Full Professor at the School of Information Systems, Singapore Management University. He received Ph.D. from the University of Minnesota, Minneapolis in 1994. His research interests include social network and web mining, information integration, and digital libraries.  He has published more than 300 papers at international journals and conferences. He currently serves as an Associate Editor of the IEEE Transactions on Knowledge and Data Engineering (TKDE), ACM Transactions on Information Systems (TOIS), ACM Transactions on the Web (TWeb), Information Processing and Management (IPM), Social Network Analysis and Mining (SNAM), and Journal of Web Engineering (JWE). 
\end{IEEEbiography}

\begin{IEEEbiography}{Freddy Chong Tat Chua}
is a Research Scientist at the Mechanisms and Design Lab of Hewlett Packard Enterprise Labs based in Palo Alto, California, USA. He is also an Adjunct Faculty in the School of Information Systems, Singapore Management University (SMU). Dr. Chua obtained Ph.D. in Information Systems from SMU in 2013, and Bachelor of Computer Science from National University of Singapore in 2007. He serves as a Program Committee member at several academic communities. He has published extensively in prominent computer science conferences and journals.
\end{IEEEbiography}

\begin{IEEEbiography}{Jia-Wei Low}
is a Data Scientist at the Infocomm Development Authority (IDA), Singapore. Prior to IDA, he worked as a Research Engineer at the Living Analytics Research Centre, School of Information Systems, Singapore Management University. He co-founded a technology startup company specializing in electricity monitoring
and home automation in 2011. His recent works include predictive modelling for public healthcare sector and emergency response.
\end{IEEEbiography}

\begin{IEEEbiography}{David Lo}
is an Assistant Professor at the School of Information Systems, Singapore Management University. He works in the area of data mining and software engineering. His primary research interests include frequent pattern mining, social network mining, dynamic program analysis, and specification mining. He received a Ph.D. in computer science from the National University of Singapore in 2008.
\end{IEEEbiography}


\end{document}

%% file: introduction.tex
\IEEEraisesectionheading{\section{Introduction}\label{sec:introduction}}

In recent years, we have witnessed a dramatic growth in social interactions taking place in social media such as Twitter and Facebook. These social media sites allow users to share contents (e.g., text, images, videos or web links), and to build social relationships, user communities, and common interest groups. Social media also generate a massive amount of digital data about user behaviors. The availability of such data has sparked the desire to learn more about consumers/users, fueling in turn the emergence of new services for peer interaction, marketing, and content sharing. For these services, there is a need to profile user preferences and attributes so as to support personalization, advertising, and recommendation \cite{Rao2010,Mislove2010}. 

Despite the abundance of user-generated data, meta data about personal attributes that are directly useful for personalized services and recommendations are often not available. In Twitter, for instance, users rarely provide demographic information as gender, age, religion, or marital status. Such information can be used by Twitter or other organizations to perform market segmentation, contextualize search engine, or make better content/friend recommendations. Recent studies \cite{Rao2010,Mislove2010,Kosinski2013,Li2014} have nonetheless shown that it is possible to use statistical means to profile the latent user attributes, based on public data (e.g., users' contents and social ties) that the users reveal in social media. 

Existing works on profiling latent user attributes in social media generally involve two types of data: content information and social connectivity \cite{Rao2010,Mislove2010,Ikeda2013}. However, two major issues hinder the widespread use of these approaches. First, most (if not all) existing methods utilize users' own contents and/or one specific type of user relationship. They are not able to fuse different types of social relationship when one's own content is unavailable (e.g., a Twitter user who has no tweets), or when one type of relationship is not sufficiently informative of the attribute of interest. Second, the current approaches employ supervised learning methods that cannot generalize well when labeled data are scarce. This necessitates a more robust method that can also exploit a large pool of unlabeled data.

\subsection{Motivating Example}

\begin{figure}[!t]
\centering
\includegraphics[width=0.45\textwidth]{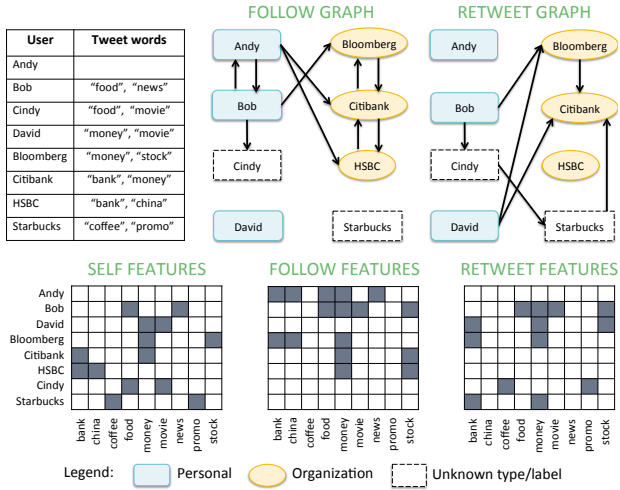}
\caption{Example of user profiling task}
\label{fig:toy_example}
\end{figure}

To illustrate these more clearly, Figure \ref{fig:toy_example} gives an example for the task of inferring if a Twitter user account is personal or organizational. The example consists of six personal and organization accounts whose labels are known (ovals and round boxes), and two accounts with unknown labels (dashed boxes). The upper half of Figure \ref{fig:toy_example} shows two types of relationships: ``follow'' and ``retweet from'', and the tweet contents (words) of each user are shown in the left table. The bottom half shows the bag-of-words feature representation of the users. Traditionally, one can infer the label using only a user's own contents (self features). Often, however, the self features alone are not indicative of the label, e.g., for user \emph{Andy} who never tweets. Augmenting social features derived from \emph{Andy}'s followees (\emph{Bob}, \emph{Citibank} and \emph{HSBC}) can provide stronger cues for Andy's label. 

In this spirit, recent works \cite{Li2014,Li2014a} have tried to incorporate social features, largely derived from a single type of relationship. But due to the sparse nature of users' connectivity, utilizing a single type of relationship may still be inadequate. For example, building a predictive model for \emph{David}'s account type can hardly benefit from social ties if we consider only his ``follow'' connections, which he has none. Hence, we are not able to gain additional word features from his followers. Similarly for \emph{Andy}, \emph{Citibank} and \emph{HSBC}, there is no additional word feature if we use only the ``retweet'' links, i.e., all of them do not have ``retweet'' links. Intuitively, integrating social features from multiple, complementary sources can boost the confidence in the label prediction. For instance, modeling \emph{Bob}'s label can benefit from the co-occurrence of the words that appear in his follow and retweet features. Finally, it is possible to build a more robust model by utilizing unlabeled data. For instance, \emph{Bob}'s shares a common word ``food'' with \emph{Cindy}'s, but has no common word with \emph{Starbucks}. Exploiting this, we can create a better classification boundary that makes \emph{Bob}'s label closer to \emph{Cindy}'s' and further away from \emph{Starbucks}.

\subsection{Proposal and Contributions}

Generalizing the above scenario, we propose a new take on user profiling task by answering several research questions: 
\begin{itemize} 
\item How can we exploit multiple types of social relationship and unlabeled data in order to infer/profile the (latent) user attributes better?
\item How do we develop an efficient and robust profiling method that can integrate multiple relationship types and unlabeled data in a computationally sound way?
\item Can we understand the contributions of different features and relationship types, as well as infer/profile on a larger user population?
\end{itemize}

In light of these questions, we present in this paper a new computational method dubbed \emph{collective semi-supervised learning} (CSL), for modeling user attributes in social media. To our best knowledge, this work is the first to formulate user profiling as the problem of jointly exploiting multiple relationship types and unlabeled data, and hence CSL provides a unified approach to solve this problem in a computationally principled and efficient manner. In particular, CSL models multiple relationship types by generically constructing \emph{multi-relational features} (MRF), and then integrates the concept of \emph{convex divergence} (CD) regularization in order to establish a convex formulation of semi-supervised learning utilizing unlabeled data. 

We summarize our main contributions as follows:
\begin{itemize}
\item We develop a simple method for collective learning via MRF, which takes into account---for different types of relationships---both the tie strength between a given user and its neighbors as well as the features of the neighbors. Deviating from existing multi-relational learning methods, which either treat multi-relational information as constraints to the learning process \cite{Wang2009,Wang2012} or rely on low-rank assumption to decompose multi-relational data \cite{Harshman1994,Singh2008,Chu2009}, our MRF approach is more general and makes less restrictive assumption about the multi-relational information.

\item We put forward the concept of CD regularization that offers a convex formulation of semi-supervised learning using unlabeled data instances (i.e., unlabeled users). This leads to a computationally sound learning procedure that warrants a unique, globally optimal solution, which can be readily identified via off-the-shelf numerical optimization methods (e.g., the Quasi-Newton algorithm in \cite{Liu1989}). This makes our CSL approach accurate, robust, and scalable. 

\item We extensively evaluate our CSL approach through two user profiling tasks in Twitter: modeling users' \emph{account type} and \emph{marital status}. The results demonstrate the accuracy and robustness of our approach, and how different relationship types and unlabeled data contribute to its performance. We also show how CSL can be used to unravel important features for different relationship types, and to infer/profile on a larger user population in Twitter.
\end{itemize}




\subsection{Paper Outline}

The remainder of this paper is organized as follows. In Section \ref{sec:related_work}, we first give an overview of related works. Section \ref{sec:framework} elaborates the proposed CSL approach. In Section \ref{sec:problem}, we describe the user profiling tasks addressed in this work, followed by the corresponding experimental results and analyses in Section \ref{sec:experiments}. Finally, we conclude in Section \ref{sec:conclusion}.

%% file: related_work.tex
\section{Related Work}
\label{sec:related_work}

We first survey related works on user attribute profiling, semi-supervised learning, and multi-relational learning. We then discuss how our approach differs from these works.

\subsection{User Attribute Profiling}
\label{sec:profiling_review}

Several works have recently been developed to infer multiple user attributes in social media. Rao \emph{et al.} \cite{Rao2010} proposed a set of network structure-based features to infer the attributes of Twitter users, including gender, age, geographical origin, and political preference. These features were then fed into a stacked classifier to infer the attributes. Mislove \emph{et al.} \cite{Mislove2010} used both global and local community detection methods in order to find communities of users who share common attribute values. Ikeda \emph{et al.} \cite{Ikeda2013} used social communities to infer the demographic information of the Twitter users. They developed a hybrid method that uses text features as well as network structure features.

Recently, Kosinski \emph{et al.} \cite{Kosinski2013} showed that public information obtained from Facebook can be used to predict demographic attributes of users. 
By factorizing a sparse matrix representing which user likes which subject/topic, a low-rank representation of the attributes was obtained and then used as features for regression and classification.
Li \emph{et al.} \cite{Li2014} devised a distant supervised learning method to infer the attributes of Twitter users by augmenting structured auxiliary data from the Facebook and Google+ networks. The unstructured information is matched with the structured ground truth to increase the prediction accuracy. 

In \cite{Li2014a}, Li \emph{et al.} developed a new co-profiling to jointly infer the users' attributes (partially observed) and relationship type (completely unobserved) within the users' ego network (observed). The underlying assumption is that social connections are discriminatively correlated with user attributes (e.g., employer, college) via a hidden relationship type (e.g., colleague, classmate).
Dong \emph{et al.} \cite{Dong2014} presented a factor graph model to predict the demographic attributes of mobile phone users. The model utilizes three types of factor: attribute factor, dyadic factor, and triadic factor, which represent correlation between the user's attributes and his/her network characteristics, between the attributes of two users, and among the attributes of user triads, respectively. 

\subsection{Semi-Supervised Learning}
\label{sec:ssl_review}

The literature on SSL methods is vast, so here we review only methods that are most related to our work. The simplest form of SSL is \emph{bootstrapping}, whereby a classifier is first trained using labeled data, and then applied to unlabeled data so as to generate more labeled samples for the next rounds of training \cite{Abney2004,Haffari2007}. Bootstrapping works based on a simplistic assumption is that the classifier's own (high-confidence) predictions are correct. \emph{Co-training} is an extension of bootstrapping in which two (or more) classifiers are trained on different, ideally disjoint sets of features, and generate labeled samples to improve each other \cite{Blum1998,Mitchell2004}. This method is less prone to mistakes than bootstrapping, but it requires that natural feature splits exist in the data.


Another class of SSL methods uses the \emph{low-density assumption} \cite{Chapelle2006}, encouraging the decision boundary to lie in low-density regions for improving generalization results. The most common way to achieve this is to use a maximum margin algorithm such as transductive support vector machine \cite{Joachims1999}. However, the corresponding learning problem is nonconvex, which is hard to solve and does not warrant globally optimal solution. Grandvalet and Bengio \cite{Grandvalet2006} devised an alternative method based on \emph{entropy regularization} (ER). This approach encourages the posterior probability to be closer to 1 or 0 through any high-density region, while the decision boundary corresponds to intermediate probability. Again, however, the resulting problem is nonconvex. 

There are also active research works on \emph{graph-based} SSL (GSSL) methods, which treat both labeled and unlabeled data as nodes in a graph and build edges between pairs of nodes weighted by their affinities (similarities) \cite{Chapelle2006}. A popular example of GSSL methods is label spreading \cite{Zhou2004}, which iteratively propagates a node's label distribution to its neighbors according to their affinity. Generalizations have been proposed under the umbrella of manifold regularization \cite{Belkin2006,Chapelle2006}. We note, however, that the GSSL methods work well only when the affinity or manifold assumption holds for the data, i.e., nodes that are similar would likely have similar label distribution. As such, the GSSL methods requires the right choice of affinity graph to work well.



Extending the entropy regularization method \cite{Grandvalet2006}, several \emph{information-theoretic} SSL methods have been developed \cite{Mann2007,Niu2013}. Mann and McCallum \cite{Mann2007} proposed the expectation regularization (XR) to build a simple and robust SSL method. The XR augments the learning procedure with a regularization term that minimizes the Kullback-Leibler (KL) divergence between the label expectations predicted by the model and human-provided label expectation priors. More recently, Niu \emph{et al.} \cite{Niu2013} devised a squared-loss mutual information regularization (SMIR) method, which led to a convex SSL problem formulation guaranteed under a mild condition. The key appeal of this approach is that an analytical (closed-form) solution can be computed to identify unique, globally optimal model parameters \cite{Suzuki2009,Niu2013}.

\subsection{Multi-Relational Learning}
\label{sec:mrl_review}

Multi-relational learning (MRL) is applicable when the data are available in multiple structured formats and can be represented as multiple graphs (a.k.a. \emph{multigraph}). That is, a multigraph can be used in MRL to encode different types of relationship (edge) among entities (nodes). 
In \cite{Xu2009}, Xu \emph{et al.} presented a seminal work on multi-relational Gaussian process (MRGP) that utilizes a generative probabilistic model based on Gaussian process. It combines the covariance and random variables approaches to model multiple relations, which in turn provides support for multiple relational learning tasks with multiple types of entities and relations. 

In a different task domain, Wang \emph{et al.} \cite{Wang2009} proposed a MRL method for video annotation that integrates multiple graphs into a regularization framework, so as to sufficiently exploit their complementation. This method was shown to be equivalent to first fusing multiple graphs and then conducting graph-based SSL on the fused graph. A similar approach was used in \cite{Wang2012} to tackle the task of protein domain ranking in structural biology. In this approach, the intrinsic manifold of protein domain distribution was approximated by combining multiple graphs for regularization. 

Another branch of relational learning considers the relations among entities as resulting from the latent factors of these entities. These approaches often translate into learning an embedding of the entities, which corresponds to a \emph{matrix factorization} problem. This can be naturally extended to MRL by stacking the matrices to be factorized and then applying tensor factorization methods \cite{Harshman1994,Chu2009}. Another natural extension to MRL is to share the common embedding or the entities across relations via collective matrix factorization \cite{Singh2008,Nickel2012}. This method has shown state-of-the-art performances on relational datasets \cite{Nickel2012}, although the number of relation types is usually modest (less than 100). Extensions have recently been proposed in \cite{Jenatton2012,Drumond2014} to handle multi-relational data with a large number of relation types.

\subsection{Our Approach}

Our CSL approach differs from the existing works in several important ways, which we elaborate below. 

\emph{Comparisons with existing user profiling methods}. While many of the current profiling methods utilize users' social information for attribute predictions, they have focused on just one type of relationship (e.g., only the follow relationship in Twitter), lacking a systematic method to incorporate different types of relationships altogether. Second, the existing profiling methods utilize only labeled data, which are often very scarce. A more robust predictive model can be obtained by also exploiting a large pool of unlabeled data. CSL offers these two capabilities in a unified and synergistic manner, which---to the best of our knowledge---is the first of its kind for user profiling applications. 

\emph{Comparisons with existing SSL methods}. In contrast to conventional SSL methods such as bootstrapping \cite{Abney2004} and co-training \cite{Blum1998}, our CSL approach does not rely on the assumption that the model's own (high-confidence) prediction is correct, or that natural feature splits exist in the data. Compared to the GSSL methods such as label spreading \cite{Zhou2004}---whose performance is sensitive to the choice of affinity graph---our approach works based on the empirical distribution of unlabeled data, which is simpler and less restrictive. Our approach also provides a convex formulation of SSL that is more robust and computationally elegant than the ER method \cite{Grandvalet2006}, whereby the learning procedure can be easily trapped to one of the (multiple) local optimal solutions. Finally, the CSL approach is more general than the state-of-the-art information-theoretic SSL methods such as XR \cite{Mann2007} and SMIR \cite{Niu2013}. These methods have not accounted for multi-relational information in their formulation.

\emph{Comparisons with existing MRL methods}. Our CSL approach compares favourably to the MRGP method \cite{Wang2009} in several ways. First, CSL adopts a discriminative probabilistic model, which should in principle be more accurate than the generative model used in MRGP \cite{Sutton2012}. Second, MRGP handles only binary relations (graphs), whereas ours can take weighted graphs. Third, MRGP is trained using the expectation maximization (EM) algorithm, which does not warrant a globally optimal solution. CSL is also less restrictive than the MRL methods in \cite{Wang2009,Wang2012}. The latter treat multi-relational information (i.e., multigraph) as constraints to the learning process, whereas ours casts multi-relational information into multi-relational features that in turn serve as additional information to be augmented into the learning process. Finally, CSL is more generic/flexible than the matrix/tensor factorization methods \cite{Harshman1994,Chu2009,Singh2008,Nickel2012,Jenatton2012,Drumond2014}. These methods rely on low-rank assumption for matrix/tensor decomposition, and do not yet cater for explicit (i.e., non-latent) features defined for each entity. 

%% file: framework.tex
\section{Proposed Framework}
\label{sec:framework}

Our CSL framework operates based on two inputs: 1) partially labeled data, comprising a feature matrix $\mathbf{X}$ with known labels $\mathbf{Y}_L$ and missing labels $\mathbf{Y}_U$, and 2) multigraph, composed of multiple directed graphs $\mathbf{G}_m$ that encode different types of social relationship. We first describe our notations: Let $\mathbf{G} = \{ \mathbf{G}_1,\ldots,\mathbf{G}_m,\ldots,\mathbf{G}_M \}$ be a multigraph composed of $M$ graphs, where each graph $\mathbf{G}_m = (\mathbf{V}, \mathbf{E}_m)$ comprises nodes $\mathbf{V}$ and edges $\mathbf{E}_m$. Note that here we have a common set of nodes $\mathbf{V}$, but  different sets of edges $\mathbf{E}_m$. We denote the feature matrix of nodes $\mathbf{V}$ as $\mathbf{X} \in \mathbb{R}^{N \times J}$, and the weighted adjacency matrix of edges $\mathbf{E}_m$ as $\mathbf{W}_m \in \mathbb{R}^{N \times N}$, where $N$ and $J$ are the number of nodes and features respectively. We also denote the labels of $\mathbf{V}$ as $\mathbf{Y}$, representing the user attributes of interest. Lastly, a node maps exactly to a data instance, so we shall use the two terms interchangeably.

\subsection{Probabilistic Foundation}
\label{sec:probability}

We first outline the probabilistic formulation of our CSL approach here. 
Ultimately, our goal is to maximize the posterior distribution of the model parameters $\Theta$, given the labels $\mathbf{Y}$, node features $\mathbf{X}$, and multigraph $\mathbf{G}$. Here the posterior can be computed using the Bayes' rule:
\begin{align}
p(\Theta | \mathbf{Y}, \mathbf{X}, \mathbf{G}) &= \frac{p(\Theta, \mathbf{Y}, \mathbf{X}, \mathbf{G})}{p(\mathbf{Y}, \mathbf{X}, \mathbf{G})} = \frac{p(\mathbf{Y} | \mathbf{X}, \mathbf{G}, \Theta) p(\Theta)}{p(\mathbf{Y}, \mathbf{X}, \mathbf{G})} \nonumber\\
&\propto p(\mathbf{Y} | \mathbf{X}, \mathbf{G}, \Theta) p(\Theta)
\end{align}


In this work, we focus on \emph{partially labeled} data, whereby only a few data instances have observed labels $\mathbf{Y}_L$, while the remaining instances are largely unlabeled, i.e., their labels $\mathbf{Y}_U$ are assumed to be \emph{missing at random} \cite{Grandvalet2006}. Since $\mathbf{Y} = \mathbf{Y}_L \cup \mathbf{Y}_U$, it follows that $p(\mathbf{Y} | \mathbf{X}, \mathbf{G}, \Theta) = p(\mathbf{Y}_L | \mathbf{X}, \mathbf{G}, \Theta) p(\mathbf{Y}_U | \mathbf{X}, \mathbf{G}, \Theta)$ and the posterior becomes:
\begin{align}
\label{eqn:posterior}
p(\Theta | \mathbf{Y}, \mathbf{X}, \mathbf{G}) &\propto p(\mathbf{Y}_L | \mathbf{X}, \mathbf{G}, \Theta) p(\mathbf{Y}_U | \mathbf{X}, \mathbf{G}, \Theta) p(\Theta)
\end{align}
where $\mathbf{Y}_L$ and $\mathbf{Y}_U$ are treated as conditionally independent. In turn, we can maximize the posterior $p(\Theta | \mathbf{Y}, \mathbf{X}, \mathbf{G})$ by minimizing its negative logarithm (a.k.a. \emph{loss function}) $\mathcal{L}$:
\begin{align}
\label{eqn:full_loss}
\mathcal{L} = -\ln(p(\mathbf{Y}_L | \mathbf{X}, \mathbf{G}, \Theta)) - \ln(p(\mathbf{Y}_U | \mathbf{X}, \mathbf{G}, \Theta)) - \ln(p(\Theta))
\end{align}
For convenience, we break (\ref{eqn:full_loss}) into two parts, respectively:
\begin{align}
\mathcal{L}_L &= -\ln(p(\mathbf{Y}_L | \mathbf{X}, \mathbf{G}, \Theta)) - \ln(p(\Theta))\\
\mathcal{L}_U &= -\ln(p(\mathbf{Y}_U | \mathbf{X}, \mathbf{G}, \Theta))
\end{align}

\textbf{Remark}. It must be noted that the above formulation is new and different from the contemporary user attribute profiling methods \cite{Rao2010,Mislove2010,Ikeda2013,Li2014,Li2014a,Dong2014}. All these approaches focus only on a single type of relationship, whereas ours can readily cater for multiple types of relationship (i.e., multigraph $\mathbf{G}$). Moreover, the existing methods do not yet exploit the additional information provided by unlabeled data (i.e., $\mathbf{Y}_U$) in guiding their learning processes. 

\subsection{Base Model}
\label{sec:logreg}

The proposed CSL approach can be viewed as a generalization of the contemporary \emph{logistic regression} model \cite{Fan2008}. Traditionally, logistic regression learns in a fully supervised fashion based solely on the labeled data $\mathbf{Y}_L$ (i.e., it does not use $\mathbf{Y}_U$), and it does not take into account multi-relational information encoded as multigraph $\mathbf{G}$. That is, by excluding $\mathbf{Y}_U$ and $\mathbf{G}$
and by assuming independent and identically distributed (i.i.d) data instances, logistic regression essentially learns to minimize the following loss function:
\begin{align}
\mathcal{L}_{L} &= -\sum_{i=1}^L \ln(p(y_i | x_i, \Theta)) - \sum_{j=1}^J \ln(p(\theta_j))
\end{align}
where $y_i \in  \mathbf{Y}_L$ and $x_i \in \mathbf{X}$ are the actual label and feature vector for data instance $i$ respectively, $L = |\mathbf{Y}_L|$ is the number of labeled data instances, and $\theta_j \in \Theta$ is a model (i.e., weight) parameter that we want to learn for each feature $j$.

Without loss of generality, we consider binary class label\footnote{Extension to multi-class task with $C > 2$ labels is straightforward, which can be done by constructing $C$ binary logistic regression models.} $y_i \in \{0, 1\}$. For binary classification, we may take that each sample likelihood $p(y_i | x_i, \Theta)$ follows a \emph{Bernouli} distribution (which is analogous to the toss of a coin): 
\begin{align}
\label{eqn:likelihood}
p(y_i | x_i, \Theta) &= \sigma_i^{y_i} (1 - \sigma_i)^{(1 - y_i)}
\end{align}
where $\sigma_i = \sigma(f(x_i, \Theta)) = \frac{1}{1 + \exp(-f(x_i, \Theta))}$ refers to the logistic function, and $f(x_i, \Theta)$ is the linear model:
\begin{align}
\label{eqn:base_model}
f(x_i, \Theta) &= \sum_{j=1}^J \theta_j x_{i,j}
\end{align}

For the prior $p(\theta_j)$, we use a Gaussian distribution  with zero mean and inverse variance $\lambda$:
\begin{align}
\label{eqn:prior}
p(\theta_j) &= \frac{1}{Z} \exp \left( -\frac{\lambda}{2} \theta_j^2 \right)
\end{align}
where $Z$ is a normalizing constant and $\lambda > 0$.
Accordingly, we can write the overall loss $\mathcal{L}_{L}$ for logistic regression as:
\begin{align}
\label{eqn:labeled_loss}
\mathcal{L}_{L} = &-\sum_{i=1}^L \left[ y_i \ln(\sigma_i) + (1 - y_i) \ln(1 - \sigma_i) \right] + \frac{\lambda}{2} \sum_{j=1}^J \theta_j^2 
\end{align}

Note that the regularization term $\frac{\lambda}{2} \sum_{j=1}^J \theta_j^2$ serves to penalize large values of the model parameters $\theta_j$, thereby reducing the risk of data overfitting \cite{Fan2008}.  



\subsection{Multi-Relational Features}
\label{sec:feature_expansion}

We now extend the base logistic regression model to incorporate the multi-relational information $\mathbf{G}$ through adding multi-relational features (MRF). Specifically, by incorporating $\mathbf{G}$ into the parameterization of the base model in (\ref{eqn:base_model}), we obtain an extended linear model $f(x_i, \mathbf{G}, \Theta)$:
\begin{align}
\label{eqn:ext_model}
f(x_i,\mathbf{G}, \Theta) &= \sum_{j=1}^J \theta_j x_{i,j} + \sum_{m=1}^M \sum_{j=1}^J \alpha_{m,j} \pi_{m,i,j}
\end{align}
and correspondingly $\sigma_i = \sigma(f(x_i,\mathbf{G}, \Theta))$, where $\pi_{m,i,j}$ is the $j^{th}$ \emph{relational feature} of data instance (i.e., node) $i$ for graph $\mathbf{G}_m$, and $\alpha_{m,j}$ is the corresponding $j^{th}$ \emph{relational weight} for $\mathbf{G}_m$, and $\Theta = \{ \theta_j \} \cup \{ \alpha_{m,j} \}$ is the set of all model parameters. Under this notation, we call $\theta_j$ as the \emph{self weight} corresponding to the \emph{self features} $x_{i,j}$ of node $i$. 

There are numerous ways to define the relational feature $\pi_{m,i,j}$ of a node $i$. In principle, one can derive the relational features through an arbitrary \emph{aggregation} function summarizing some global or local properties of each graph $\mathbf{G}_m$, and the aggregation function need not be the same for different graphs $\mathbf{G}_m$. For efficiency and interpretability, however, in this work we focus on a simple aggregation function that combines the information from only the \emph{immediate neighbors} of a node $i$ by taking a weighted average of their features:
\begin{align}
\label{eqn:augment}
\pi_{m,i,j} &= \frac{\sum_{(i,i') \in \mathbf{E}_m} w_{m,i,i'} x_{i',j}}{\sum_{(i,i') \in \mathbf{E}_m} w_{m,i,i'}} 
\end{align}
where $w_{m,i,i'} \in \mathbf{W}_m$ represents the tie strength of a node (instance) $i$ with its neighbor $i'$ in graph $\mathbf{G}_m$. For instance, in the context of Twitter follow graph, the notion of neighbors refers to the followees of a given user.

With the addition of the relational weights $\alpha_{m,j}$, the penalized loss $\mathcal{L}_L$ now becomes:
\begin{align}
\label{eqn:mrf_loss}
\mathcal{L}_L = &-\sum_{i=1}^L \left[ y_{i} \ln(\sigma_{i}) +  (1 - y_{i}) \ln(1 - \sigma_{i}) \right] \nonumber\\
                &+ \frac{\lambda}{2} \sum_{j=1}^J \left( \theta_j^2 + \sum_{m=1}^M \alpha_{m,j}^2 \right)
\end{align} 

\textbf{Remark}. The MRF formulation in (\ref{eqn:ext_model}) and (\ref{eqn:augment}) provides a simple yet powerful way to incorporate multiple types of social relationship into user attribute prediction. Such formulation has several key appeals:

\begin{itemize}
\item Unlike previous MRL methods that use multigraph to \emph{constrain} the learning processes, e.g., \cite{Wang2009,Wang2012}, or rely only on \emph{latent features}, e.g., \cite{Harshman1994,Chu2009,Nickel2012}, our MRF formulation is more generic and makes less stringent assumption. That is, we treat multigraph as \emph{additional source} of information, and we can use any aggregation function to summarize this information. 

\item Our MRF formulation can also readily cater for different types of features, such as numeric features (e.g., tweet count), $n$-gram representation of text features, or binary vector of categorical features. For ease of interpretation/analysis, though, we shall focus on the $n$-gram text features in this work.

\item By aggregating and augmenting the neighbors' features on a per-graph basis, we can exploit the dependencies and complementarity among various features, while preserving the semantics of each type of relationship. Especially, the learned relational weights $\alpha_{m,j}$ can be used to understand the contribution and importance of different types of relationship in modeling latent user attributes. 

\item From a computational standpoint, the MRF formulation is efficient. First, the aggregation function keeps the problem dimensionality moderate; we only require $(M + 1) \times J$ features instead of na\"{i}vely appending all neighbors' features. Second, the relational features $\pi_{m,i,j}$ can be pre-computed once for every instance/node $i$ prior to parameter learning process. Finally, our MRF formulation maintains the linearity of our model (\ref{eqn:ext_model}), which preserves the convexity of the overall loss (\ref{eqn:full_loss}) (see Section \ref{sec:optimization}).
\end{itemize}

It is also worth noting that our MRF formulation is different from that of conditional random field (CRF) \cite{Sutton2012}. The CRF approach usually involves some form of dependencies among the (output) labels $y_i$, whereas our MRF method focuses on the dependencies in the input space $x_i$ and assumes that the labels $y_i$ are (conditionally) independent. While structured modeling via CRF can potentially improve performance, it comes at the expense of higher computational complexity and degraded model interpretability. As such, we do not pursue the CRF approach in this work.

\subsection{Convex Divergence Regularization}
\label{sec:convex_divergence}

After constructing the MRF for all data instances (both labeled and unlabeled), CSL carries out a semi-supervised learning (SSL) using unlabeled data for improving model generalization and robustness. To this end, we put forward the idea of \emph{convex divergence} (CD) to regularize learning via unlabeled data. The CD regularization stems from the following definition of $p(\mathbf{Y}_U | \mathbf{X}, \mathbf{G}, \Theta)$ in (\ref{eqn:posterior}): 
\begin{align}
\label{eqn:unlabeled_dist}
p(\mathbf{Y}_U | \mathbf{X}, \mathbf{G}, \Theta) &= \frac{1}{Z} \exp \left(-\beta D_f(\mu || \rho) \right)
\end{align}
where $Z$ is a normalizing constant, $\beta$ is a (nonnegative) user-specified regularization parameter, and $D_f(\mu || \rho)$ is the $f$-divergence \cite{Ali1966} between two distributions $\mu$ and $\rho$:
\begin{align}
D_f(\mu || \rho) &= \sum_z f\left( \frac{\mu(z)}{\rho(z)} \right) \rho(z)  
\end{align}
which is defined over some space $z$ and $f: [0, +\infty) \rightarrow \mathbb{R}^+$ is a continuous convex function, such that $f(1) = 0$. 

We note here that $D_f(\mu || \rho)$ does not uniquely define the form of the prior distribution $p(\mathbf{Y}_U | \mathbf{X}, \mathbf{G}, \Theta)$, but the latter can be constructed through constraints imposed by $D_f(\mu || \rho)$. Also, to ensure convexity in the overall loss $\mathcal{L}$, it is necessary to choose the distributions $\mu$ and $\rho$ such that $D_f(\mu || \rho)$---or its approximation---is twice-differentiable, and its second derivative $\bigtriangledown^2 D_f(\mu || \rho)$ is nonnegative for all possible values of $z$. 

In this work, we focus on an instantiation of $D_f(\mu || \rho)$ that involves computing the \emph{Kullback-Leibler} (KL) divergence between some class prior $\tilde{p}$ and the expected predictions $\mathbb{E}[\sigma_{i}]$ made by the model on unlabeled data:
\begin{align}
D_{KL}(\mu || \rho) &= \tilde{p} \ln \left( \frac{\tilde{p}}{\mathbb{E}[\sigma_{i}]} \right) \nonumber\\
&= -\tilde{p} \ln(\mathbb{E}[\sigma_{i}]) + \underbrace{\tilde{p} \ln \left( \tilde{p} \right)}_{= constant} \nonumber\\
\label{eqn:kl_divergence}
&\propto -\tilde{p} \ln(\mathbb{E}[\sigma_{i}])
\end{align} 
where the function $f$ is defined as $f(t) = t \ln(t)$. 
In this case, $D_{KL}(\mu || \rho) = 0$ when $\mu$ and $\rho$ match exactly. Our goal here is to \emph{minimize} (\ref{eqn:kl_divergence}), implying that we want to obtain a classification model such that the expectation of its predictions on unlabeled data is similar to the class prior. 

The class prior can be either (manually) specified based on domain knowledge, or computed based on the class distribution on the labeled data. For simplicity, we choose the latter approach in this work, by defining $\tilde{p}$ as $\tilde{p} = \frac{1}{L} \sum_{i=1}^L y_{i}$,
i.e., the proportion of positive instances in the labeled data.

Finally, by subtituting (\ref{eqn:kl_divergence}) into (\ref{eqn:unlabeled_dist}) and dropping constant terms, we obtain the CD regularization:
\begin{align}
\label{eqn:unlabeled_loss}
\mathcal{L}_U = &-\ln \left( p(\mathbf{Y}_U | \mathbf{X}, \mathbf{G}, \Theta) \right) \nonumber\\
              \propto &- \ln \left(- \beta D_{KL}(\tilde{p} || \hat{p}_{\Theta}) \right) \nonumber\\
              \propto &-\beta \text{ } \tilde{p} \ln \left( \mathbb{E}[\sigma_{i}] \right)
\end{align}
Note that (\ref{eqn:unlabeled_loss}) is not convex in its current form. Fortunately, we can use the \emph{Jensen inequality} \cite{Jensen1906} in order to derive a convex upper bound of (\ref{eqn:unlabeled_loss}). The Jensen inequality states that the expectation of a convex function $\varphi$ is equal to or greater than the function of the expectation, i.e., $\mathbb{E}[\varphi(x)] \geq \varphi(\mathbb{E}[x])$. It then follows that the convex upper bound is:
\begin{align}
\mathcal{L}_U \propto \text{ } &\beta \text{ } \tilde{p} \left( -\ln \left( \mathbb{E}[\sigma_{i}] \right) \right) \nonumber\\
              \leq \text{ } &\beta \text{ } \tilde{p} \text{ }\mathbb{E} \left[ -\ln \left( \sigma_{i} \right) \right]
\end{align}
with $\varphi(x) = -\ln(x)$. Subsequently, we can approximate the expectation via an empirical average $\mathbb{E} \left[ -\ln \left( \sigma_{i} \right) \right] = -\frac{1}{U} \sum_{i=L+1}^{L+U} \ln \left( \sigma_{i}) \right)$, where $U = |\mathbf{Y}_U|$ is the total number of unlabeled data instances. This leads to a new convex formulation of $\mathcal{L}_U$ using unlabeled data:
\begin{align}
\label{eqn:convex_loss}
\mathcal{L}_U \propto &-\beta \text{ } \tilde{p} \sum_{i=L+1}^{L+U} \ln \left( \sigma_{i} \right) 
\end{align}
whereby we absorb the term $\frac{1}{U}$ into $\beta$ for simplicity.

\textbf{Remark}. The formulation in (\ref{eqn:convex_loss}) is related to the XR approach \cite{Mann2007}, with some key differences. First, the XR method tries to minimize (\ref{eqn:kl_divergence}) directly, which is non-convex and may lead to local optima. In contrast, our CD formulation aims at reducing (\ref{eqn:kl_divergence}) by minimizing its convex upper bound (\ref{eqn:convex_loss}), which is simpler and computationally more appealing (due to convexity). Second, our formulation generalizes XR by not only learning from unlabeled data, but also taking into account the different types of relationship among instances via MRF. We will empirically show in Section \ref{sec:experiments} how multi-relational information and unlabeled data can work together to improve user profiling performances. Our formulation is also conceptually superior to that of the SMIR method \cite{Niu2013}, whose convexity is not guaranteed when the L2 regularization parameter ($\lambda$ in our notation) is not sufficiently large \cite{Niu2013}. We will show in Section \ref{sec:optimization} that our CSL formulation imposes strict convexity for any positive $\lambda$.

\subsection{Parameter Learning}
\label{sec:optimization}

We can now combine (\ref{eqn:unlabeled_loss}), (\ref{eqn:mrf_loss}) and (\ref{eqn:ext_model}) to derive the overall loss function $\mathcal{L}$ for the CSL approach, which is given by:
\begin{align}
\label{eqn:final_loss}
\mathcal{L} = &-\sum_{i=1}^L \left[ y_{i} \ln(\sigma_{i}) +  (1 - y_{i}) \ln(1 - \sigma_{i}) \right] \nonumber\\
                &+ \frac{\lambda}{2} \sum_{j=1}^J \left( \theta_j^2 + \sum_{m=1}^M \alpha_{m,j}^2 \right) -\beta \text{ } \tilde{p} \sum_{i=L+1}^{L+U} \ln \left( \sigma_{i} \right) 
\end{align}
with $\sigma_{i,1} = \sigma(f(x_i, \mathbf{G}, \Theta))$ and $\sigma_{i,2} = 1 - \sigma(f(x_i, \mathbf{G}, \Theta))$.

\textbf{Convexity}. Before we proceed with the learning procedure for minimizing $\mathcal{L}$, we first give a proof sketch for the convexity of $\mathcal{L}$. This is done by examining the \emph{slope} (i.e., first derivative) and \emph{curvature} (i.e., second derivative) of $\mathcal{L}$. Firstly, the slope with respect to each self weight $\theta_j$ is:
\begin{align}
\frac{\partial \mathcal{L}}{\partial \theta_j} = &-\sum_{i=1}^L \frac{y_{i}}{\sigma_{i}} \frac{\partial \sigma_{i}}{\partial \theta_j} + \lambda \theta_j - \beta \text{ } \tilde{p} \sum_{i=L+1}^{L+U} \frac{1}{\sigma_{i}} \frac{\partial \sigma_{i}}{\partial \theta_j}
\end{align}
and it is easy to show that $\frac{\partial \sigma_{i}}{\partial \theta_j} = \sigma_{i}  \left( 1 - \sigma_{i} \right) x_{i,j}$,
where $\sigma_i = \sigma(f(x_i, \mathbf{G}, \Theta))$. We can then evaluate the slope as:
\begin{align}
\label{eqn:slope}
\frac{\partial \mathcal{L}}{\partial \theta_j} = &-\sum_{i=1}^L \left[ \left( \frac{y_{i}}{\sigma_{i}} - \frac{1 - y_{i}}{1 - \sigma_{i}}  \right) \frac{\partial \sigma_{i}}{\partial \theta_j} \right] + \lambda \theta_j - \beta \text{ } \tilde{p} \sum_{i=L+1}^{L+U} \frac{1}{\sigma_{i}} \frac{\partial \sigma_{i}}{\partial \theta_j} \nonumber\\
= &\sum_{i=1}^L \left[ \left( \sigma_{i} - y_{i} \right) x_{i,j} \right] + \lambda \theta_j - \beta \text{ } \tilde{p} \sum_{i=L+1}^{L+U} \left[ (1 - \sigma_{i}) x_{i,j} \right]
\end{align}

Finally, we may differentiate (\ref{eqn:slope}) to obtain the curvature:
\begin{align}
\label{eqn:curvature}
\frac{\partial^2 \mathcal{L}}{\partial \theta_j^2} = &\sum_{i=1}^L \left[ \sigma_{i} \left( 1 - \sigma_{i} \right) x_{i,j}^2 \right] + \lambda + \beta \text{ } \tilde{p} \sum_{i=L+1}^{L+U} \left[ \sigma_{i} \left( 1 - \sigma_{i} \right) x_{i,j}^2 \right]
\end{align}
It is clear that the curvature (\ref{eqn:curvature}) will always be positive for any positive $\lambda$ (since $\sigma_{i} \in [0, 1]$). We can thus conclude that the full loss $\mathcal{L}$ is \emph{strictly convex} for $\lambda > 0$. The convexity for the relational weight $\alpha_{m,j}$ can be proven in the same manner, and thus we omit the details here for brevity. 

\textbf{Algorithm}. Thanks to the convexity trait, we can use any off-the-shelf gradient-based algorithm to learn the parameters of our CSL model. In this work, we use the \emph{limited memory Broyden--Fletcher--Goldfarb--Shanno} (L-BFGS) algorithm \cite{Liu1989}, a popular, efficient Quasi-Newton method for solving unconstrained optimization problems\footnote{More specifically, we use an implementation of the L-BFGS algorithm provided in the SciPy library: http://goo.gl/q2dfnZ}. Algorithm 1 outlines the overall CSL learning procedure, combining the MRF and CD formulations. As for the stopping criterion, we terminate the algorithm when a maximum iteration $I$ (default: $10$) is reached, or the maximum projected slope is below a threshold $\epsilon$ (default: $10^{-5}$). Notably, the memory/time complexity of the L-BFGS algorithm is linear in the problem size \cite{Liu1989}, and the convexity of our CSL formulation makes it possible to reach the optimum within a few iterations.

\begin{algorithm}[!t]
\begin{algorithmic}[1]
\REQUIRE Feature matrix $\mathbf{X}$, actual labels $\mathbf{Y}_L$, multigraph $\mathbf{G}$ 
\ENSURE Model parameters $\Theta = \{ \theta_{j} \} \cup \{ \alpha_{m,j} \}$
\STATE Initialize all parameters $\theta_j$ and $\alpha_{m,j}$ to zero
\FOR {each graph $m \in \{1,\ldots,M\}$}
	\STATE Construct the relational features $\pi_{m,i,j}$ using (\ref{eqn:augment})
\ENDFOR
\REPEAT
    \STATE Generate the prediction scores $\sigma_i$ based on (\ref{eqn:ext_model})
    \STATE Compute the overall loss $\mathcal{L}$ using (\ref{eqn:final_loss})
    \STATE Compute the slopes $\frac{\partial \mathcal{L}}{\partial \theta_j}$ (as well as $\frac{\partial \mathcal{L}}{\partial \alpha_{m,j}}$) via (\ref{eqn:slope})
    \STATE Perform an L-BFGS iteration using $\mathcal{L}$ and $\frac{\partial \mathcal{L}}{\partial \theta_j}$ $\left( \frac{\partial \mathcal{L}}{\partial \alpha_{m,j}} \right)$
\UNTIL stopping criterion is met
\end{algorithmic}
\caption{CSL Learning Procedure}
\label{algo:incrementalEM}
\end{algorithm}




%% file: problem.tex
\section{User Profiling in Twitter}
\label{sec:problem}

This section provides an overview of the Twitter dataset and user profiling tasks we consider in this work. 

\subsection{Twitter Dataset}
\label{sec:dataset}

In our study, we use the Twitter data of Singapore users---hereafter called \emph{SGTwitter}--collected from the period of 1--28 February 2014. Starting from a set of seed Singapore users, we crawled their network based on the follow, retweet, and user mention links. Next, we added to our user base those followers/followees, retweet sources, and mentioned users who declare Singapore as their profile location. Accordingly, we obtained a total of 130,142 public user accounts whose profiles can be accessed/studied. We then focused on active users who tweeted at least twice within 1 month, which gave us the final set of 100,497 active users.


Table \ref{tab:count_dist} summarizes the count statistics of our SGTwitter data for different activities, aggregated at the user level. In general, we can see that the activity counts follow a heavy-tail distribution. As expected, other than celebrity users, a user generally follows more users than being followed. Intuitively, a user could select followees he/she is interested in, but not the followers. Hence, we can expect the followee links to be a better representation of user interests than the follower links. On the other hand, we can see that user mention and retweet activities are much more focused/targeted, resulting in sparser connectivity in the mention and retweet graphs than in the follow graph.  We shall focus on the followee, mention and retweet links in our studies later. 



\begin{table}[!t]
\scriptsize
\centering
\caption{Count statistics of our SGTwitter data}
\begin{tabular}{l|c|c|c|c|c}
\hline
Entity      & Mean & Median & $25\%$ & $75\%$ & Maximum\\
\hline
\#tweets    & 168.45 & 48  & 12  & 173 & 16,888\\
\#followees & 349.99 & 198 & 109 & 342 & 512,978\\
\#followers & 852.32 & 174 & 80  & 333 & 4,062,786\\
\#mentions  & 123.41 & 26  & 4   & 117 & 37,457\\
\#retweets  & 54.12  & 7   & 0   & 39  & 16,860\\
\hline
\multicolumn{6}{l}{$25\%: 25^{th}$ percentile, $75\%: 75^{th}$ percentile}
\end{tabular}
\label{tab:count_dist}
\end{table}

\begin{table}[!t]
\scriptsize
\centering
\caption{Label distribution of SGTwitter data}
\begin{tabular}{l|c|l|c}
\hline
\multicolumn{2}{c|}{Account type} & \multicolumn{2}{c}{Marital status}\\
\hline
Personal & 794 		& Single & 1,304\\
Organization & 514  & Married & 1,009\\
Unlabeled & 99,189  & Unlabeled & 98,184\\
\hline
\end{tabular}
\label{tab:label_dist}
\end{table}

\subsection{User Attributes}
\label{sec:attribute}

In this work, we consider the task of classifying two user attributes (i.e., labels): \emph{account type} (i.e., \emph{personal} vs. \emph{organization}) and \emph{marital status} (i.e., \emph{single} vs. \emph{married})\footnote{While our work currently focuses on two user attributes, we note that CSL is general and readily applicable to any attributes/labels.}. Profiling these attributes is a relatively new problem that has not been well studied before. This could bring about benefits in terms of providing customized services/supports that cater for the different needs of each user type. For example, organization accounts may require a service to standardize the format of their content postings or to track sentiments on their products, whereas personal accounts would likely benefit from personalized friend and content recommendation. Similarly, married users would likely be more interested in family-related products or contents than single users. 

To derive the account type and marital status labels, we first defined several keywords/phrases describing the respective labels. For the account type task, we detected organization accounts by checking if the URLs in their profile description end with ``com.sg'', ``edu.sg'' and ``gov.sg''. We randomly sampled the remaining Twitter users, and then manually labeled and judged if they are personal accounts. For the marital status, examples of relevant keywords/phrases are ``wife'', ``spouse'', ``my son'' for married users, and ``girlfriend'', ``in a relationship'', ``my gf'' for single (unmarried) users. After identifying accounts with the relevant keywords/phrases in their profile description, we manually verified the label assignment of each account.

Table \ref{tab:label_dist} summarizes the label distribution of our SGTwitter data. Here the minority class labels are ``organization'' and ``married'' for the account type and marital status tasks, respectively. Our main interest is to correctly predict these minority cases, which are expected to form a small portion in the complete data, and are thus harder to predict.

\subsection{Feature Extraction}
\label{sec:feat_extraction}

Our primary interest here is to investigate to what extent the contents generated by a user can be used to infer his/her (latent) attributes. As such, this work shall be focused primarily on \emph{text features} derived from users' tweets, though we note that our CSL approach is generic and can work on any type of features. We use the term \emph{document} to refer to a data instance $i$, which represents the collection of tweets posted by a user $i$. In this context, our goal is to infer a user's attribute based on his/her tweet document.

To extract the text features, we first converted the raw tweets into a \emph{bag-of-words} vector, from which we can derive an $n$-gram representation suitable for our CSL model. 
We summarize our feature extraction steps as follows:

\begin{itemize}
\item \emph{Tokenization}: We broke a tweet document into its constituent word tokens, 
and then created bags of word tokens, where each bag has the frequency of the tokens appearing in a document. Prior to tokenization, we also converted all letters to lowercase and devised regular expressions to extract and retain special entities such as emoticons, HTML/URL tags, phone numbers, and hashtags.  

\item \emph{Stop-word removal}: We then omitted words that appear very frequently and contribute little to discriminating the tweets of a user from those of another user. We used the list of English stop-words in \cite{Lewis2004}.

\item \emph{Normalization}: To normalize the word frequencies, we applied the \emph{term frequency--inverse document frequency} (TF-IDF) scheme \cite{Baeza-Yates1999}, which puts greater importance on words that appear frequently in a document, and deems words that occur in many documents as less important. Our TF-IDF vectors comprise $1$-gram, $2$-gram, and $3$-gram  representations \cite{Baeza-Yates1999}.   
\end{itemize}

\subsection{Multi-Relational Information}

In our study on the SGTwitter data, we consider the multi-relational information $\mathbf{G}$ derived from three directed graphs: the \emph{follow}, \emph{mention}, and \emph{retweet} graphs. The follow graph contains binary edge weights. That is, $w_{m,i,i'} = 1$ if a user $i$ follows another user $i'$ and $0$ otherwise. On the other hand, the weights of the mention/retweet graph refer to the number of times (count) of a user $i$ mentioning/retweeting user $i'$. In this case, no edge is constructed for a zero count. For each user, we consider his/her \emph{out-edges} in order to compute the relational features $\pi_{m,i,j}$ for all three graphs.

%% file: experiment.tex
\section{Experimental Results and Analysis}
\label{sec:experiments}

This section presents the results of our study on profiling the account type and marital status of the SGTwitter users. We aim at addressing several research questions (RQs):
\begin{itemize}
\item \emph{RQ1}: How does the performance of CSL compare with that of other SSL methods?
\item \emph{RQ2}: How do multi-relational features and unlabeled data contribute to the performance?
\item \emph{RQ3}: What are the important features and relationship types for predicting user labels?
\item \emph{RQ4}: How well can the learned CSL model generalize to unseen (unlabeled) data?
\item \emph{RQ5}: What can predictions made by CSL tell about a larger user population?
\end{itemize}

\textbf{Procedure}. To address the above RQs, we consider two scenarios: evaluation using labeled data (for \emph{RQ1}--\emph{RQ3}), and evaluation using unlabeled data (for \emph{RQ4}--\emph{RQ5}). For the first scenario, we adopt a \emph{stratified} 10-fold CV procedure, i.e., we split the SGTwitter data into 10 sets of training and testing data, each retaining the class label proportion as per the original data. We then report the averaged performance. Also, for all SSL methods considered in this study, their training for each fold involves using labeled instances in the training set, plus \emph{all} the remaining unlabeled instances from the full original data. For the second scenario, we manually inspect the top $k$-predicted users for each class label. The goal is to see how well our method predicts on completely unseen data (i.e., not labeled apriori), and compare the predictions with the labeled dataset.

\textbf{Metric}. To quantify performance, we examine the averaged $\text{\emph{F1-score}}$, an evaluation metric that is popularly used in text classification and information retrieval \cite{Baeza-Yates1999}. The $\text{\emph{F1-score}}$ measures classification accuracy in terms of a harmonic mean of $Precision$ and $Recall$, i.e., $\text{\emph{F1}} = \frac{2 \times Precision \times Recall}{Precision + Recall}$, where $Precision = \frac{TP}{TP + FP}$, $Recall = \frac{TP}{TP+FN}$, and $TP$, $FP$ and $FN$ are the true positives, false positives, and false negatives respectively. For these metrics, again the positive class refers to the minority labels, i.e., ``organization'' for account type and ``married''  for marital status. Lastly, we also look at the averaged \emph{training time} (in seconds) of a given method, so as to gauge its computational efficiency.

\textbf{Baseline}. We compare our CSL method with several representative SSL baseline methods. The first baseline is \emph{bootstrapping} \cite{Haffari2007}, where we first train a logistic regression using labeled data, apply it to predict on unlabeled data, and then add into the labeled dataset those samples that have high prediction scores. We repeat this for $10$ iterations, where for each iteration $n \in \{1,\ldots,10\}$, we add the top $\frac{n}{10}$ predicted samples into the labeled set. The second baseline is \emph{label spreading} (LS), a popular graph-based SSL method \cite{Zhou2004}.  We reiterate from Section \ref{sec:ssl_review} that LS relies on the affinity assumption, and its success depends on the choice of affinity matrix. We explore two renowned variants of kernel functions to define the affinity matrix in LS: $k$-nearest-neighbor ($k$NN) kernel, and radial basis (RBF) kernel \cite{Zhou2004}. 

Additionally, we compare our CSL approach with two state-of-the-art information theoretic SSL methods, namely \emph{entropy regularization} (ER) \cite{Grandvalet2006} and \emph{expectation regularization} (XR) \cite{Mann2007}. 
The overall loss functions $\mathcal{L}$ for the ER and XR methods are respectively as follows:
\begin{align}
\text{ER}: \mathcal{L} = &-\sum_{i=1}^L \left[ y_{i} \ln(\sigma_{i}) +  (1 - y_{i}) \ln(1 - \sigma_{i}) \right] \nonumber\\
                         &-\beta \sum_{i=L+1}^{L+U} \sigma_{i} \ln \left( \sigma_{i} \right)
\end{align}
\begin{align}
\text{XR}: \mathcal{L} = &-\sum_{i=1}^L \left[ y_{i} \ln(\sigma_{i}) +  (1 - y_{i}) \ln(1 - \sigma_{i}) \right] \nonumber\\
						 &+ \frac{\lambda}{2} \sum_{j=1}^J \left( \theta_j^2 + \sum_{m=1}^M \alpha_{m,j}^2 \right) - \beta \ln \left( \tilde{p} \sum_{i=L+1}^{L+U} \sigma_{i}^2 \right)
\end{align}

\textbf{Configuration}. For the ER, XR, and CSL methods, we chose the best SSL regularization parameter $\beta$ from the following candidate list: $\{10^{-4}, 10^{-3}, 10^{-2}, 10^{-1}, 1, 10, 100 \}$. Meanwhile, the L2 regularization parameter for XR and CSL was fixed to $\lambda = 1$, which we found to give good results overall. We also note that all experiments presented in this paper were carried out on a computer server with the following virtual machine configuration: 7-core Intel Xeon 2.6 GHz processor with a total of 70 GB memory (RAM).

\textbf{Significance test}. To evaluate whether the performance difference between two methods is statistically significant, we perform the \emph{Wilcoxon signed-rank test} \cite{Wilcoxon1945} with a critical value of $0.05$. The Wilcoxon test provides a non-parametric alternative to the t-test for matched pairs, when the pairs cannot be assumed to be normally distributed. When the test yields a $p$-value less than $0.05$, we deem that the performance difference is significant (and vice versa).

\subsection{Performance Comparisons (RQ1)}
\label{sec:performance}

We first evaluate the 10-fold CV results of our method using the ``full'' training set. Specifically, for each fold, we train our CSL method using $90\%$ of the available labeled data (i.e., $90\%$ of 1,308 and 2,313 labeled accounts for account type and marital status tasks respectively; cf. Table \ref{tab:label_dist}), plus all the remaining unlabeled data. Tables \ref{tab:account_benchmark} and \ref{tab:marital_benchmark} show the results for the two tasks respectively, comparing the $\text{\emph{F1-scores}}$ and training time of our CSL method with those of all the other SSL methods. To facilitate comprehensive evaluations, we present the results for different MRF settings: no graph, single graph, and all three graphs.  In addition, we show the $p$-values of the Wilcoxon test for comparing the $\text{\emph{F1-scores}}$ of the respective MRF settings (e.g., the $p$-value for bootstrapping with \emph{follow graph} features involves comparison to CSL with \emph{follow graph} features as well). As an additional reference, we include the $\text{\emph{F1-score}}$ produced by a random guess\footnote{Random guess refers to the case where $\frac{TP}{TP + FN} = \frac{FP}{FP + TN}$.} 

\begin{table}[!t]
\scriptsize
\centering
\caption{Classification results for account type classification}
\begin{tabular}{l|l|c|c|c}
\hline
Method	& Graph	& F1-Score	& Time (sec) & $p$-value\\
\hline
Random guess baseline                  & None    & 0.3930   & $-$		& 0.0025$*$\\
\hline
\multirow{5}{*}{Bootstrapping}         & None    & 0.8275	& 29.4142	& 0.0035$*$\\
									   & Follow	 & 0.8441	& 76.1515	& 0.0297$*$\\
									   & Mention & 0.8273	& 56.0481	& 0.0109$*$\\
									   & Retweet & 0.8264	& 70.8066	& 0.0142$*$\\
									   & All	 & 0.8373	& 113.2009	& 0.0047$*$\\
\hline
\multirow{5}{*}{Label spreading (kNN)} & None    & 0.5311	& 30.7675	& 0.0025$*$\\
									   & Follow	 & 0.5588	& 60.4088	& 0.0025$*$\\
									   & Mention & 0.5548	& 57.5618	& 0.0025$*$\\
									   & Retweet & 0.5282	& 58.8224	& 0.0025$*$\\
									   & All	 & 0.5477	& 114.2311	& 0.0025$*$\\
\hline
\multirow{5}{*}{Label spreading (RBF)} & None    & 0.6572	& 5.4188	& 0.0025$*$\\
									   & Follow	 & 0.7503	& 6.6155	& 0.0025$*$\\
									   & Mention & 0.6496	& 6.4276	& 0.0025$*$\\
									   & Retweet & 0.6504	& 6.4941	& 0.0025$*$\\
									   & All	 & 0.7318	& 8.9634	& 0.0025$*$\\
\hline
\multirow{5}{*}{Entropy regularization} & None   & 0.8230	& 1.0824	& 0.0372$*$\\
									   & Follow	 & 0.8481	& 4.2607	& 0.0372$*$\\
									   & Mention & 0.8169	& 3.1640	& 0.0083$*$\\
									   & Retweet & 0.8294	& 2.2800	& 0.0463$*$\\
									   & All 	 & 0.8494	& 7.8640	& 0.1013\\
\hline
\multirow{5}{*}{Expectation regularization} & None		 & 0.8490	& 0.8710 & 0.1587\\
									   & Follow	 & 0.8643	& 3.6370	& 0.1587\\
									   & Mention & 0.8454	& 3.1392	& 0.0398$*$\\
									   & Retweet & 0.8438	& 2.6258	& 0.1587\\
									   & All	 & 0.8563	& 7.9321	& 0.0544\\
\hline
\multirow{5}{*}{Our approach (CSL)}    & None    & 0.8481   & 0.9972 	& $-$\\
									   & Follow	 & 0.8635	& 4.4847 	& $-$\\
									   & Mention & 0.8491	& 3.1003 	& $-$\\
									   & Retweet & 0.8453	& 2.6968	& $-$\\
									   & All	 & 0.8608	& 8.8409	& $-$\\
\hline
\multicolumn{5}{l}{$*$: significant at a critical value of $0.05$; $-$: not applicable}
\end{tabular}
\label{tab:account_benchmark}
\end{table}

\begin{table}[!t]
\scriptsize
\centering
\caption{Classification results for marital status classification}
\begin{tabular}{l|l|c|c|c}
\hline
Method	& Graph	& F1-Score	& Time (sec) & $p$-value\\
\hline
Random guess baseline                  & None    & 0.4362   & $-$		& 0.0025$*$\\
\hline
\multirow{5}{*}{Bootstrapping}         & None 	 & 0.5420	& 20.4187	& 0.0025$*$\\
									   & Follow	 & 0.5824	& 171.1721	& 0.0025$*$\\
									   & Mention & 0.5842	& 55.4378	& 0.0025$*$\\
									   & Retweet & 0.5554	& 48.2193	& 0.0025$*$\\
									   & All	 & 0.6278	& 356.1654	& 0.0035$*$\\
\hline
\multirow{5}{*}{Label spreading (kNN)} & None    & 0.5673	& 35.5445	& 0.0035$*$\\
									   & Follow	 & 0.4575   & 66.9595	& 0.0025$*$\\
									   & Mention & 0.5625	& 65.0869	& 0.0035$*$\\
									   & Retweet & 0.5740	& 65.8956	& 0.0035$*$\\
									   & All	 & 0.4727	& 133.1662	& 0.0025$*$\\
\hline
\multirow{5}{*}{Label spreading (RBF)} & None    & 0.5215	& 5.3937	& 0.0025$*$\\
									   & Follow	 & 0.1822	& 6.6723	& 0.0025$*$\\
									   & Mention & 0.5509	& 6.8228	& 0.0035$*$\\
									   & Retweet & 0.5604	& 6.5979	& 0.0025$*$\\
									   & All	 & 0.2550	& 9.2841	& 0.0025$*$\\
\hline
\multirow{5}{*}{Entropy regularization}& None    & 0.5757	& 1.0106	& 0.0025$*$\\
									   & Follow	 & 0.6200	& 4.4185	& 0.0025$*$\\
									   & Mention & 0.5906	& 3.4064	& 0.0025$*$\\
									   & Retweet & 0.5833	& 2.3030	& 0.0025$*$\\
									   & All 	 & 0.6225	& 8.1122	& 0.0025$*$\\
\hline
\multirow{5}{*}{Expectation regularization} & None	 & 0.5821	& 1.0399	& 0.0025$*$\\
									   & Follow	 & 0.6533	& 4.1410	& 0.1206\\
									   & Mention & 0.6032	& 2.7075	& 0.0083$*$\\
									   & Retweet & 0.5966	& 2.4442	& 0.0035$*$\\
									   & All	 & 0.6513	& 7.7271	& 0.0463$*$\\
\hline
\multirow{5}{*}{Our approach (CSL)}    & None    & 0.6310	& 1.0050 	& $-$\\
									   & Follow	 & 0.6644	& 4.4516 	& $-$\\
									   & Mention & 0.6265	& 3.1527 	& $-$\\
									   & Retweet & 0.6318	& 2.4739 	& $-$\\
									   & All	 & 0.6631	& 9.6041 	& $-$\\
\hline
\multicolumn{5}{l}{$*$: significant at a critical value of $0.05$; $-$: not applicable}
\end{tabular}
\label{tab:marital_benchmark}
\end{table}

\begin{figure*}[!t]
\centering
\includegraphics[width=0.96\textwidth]{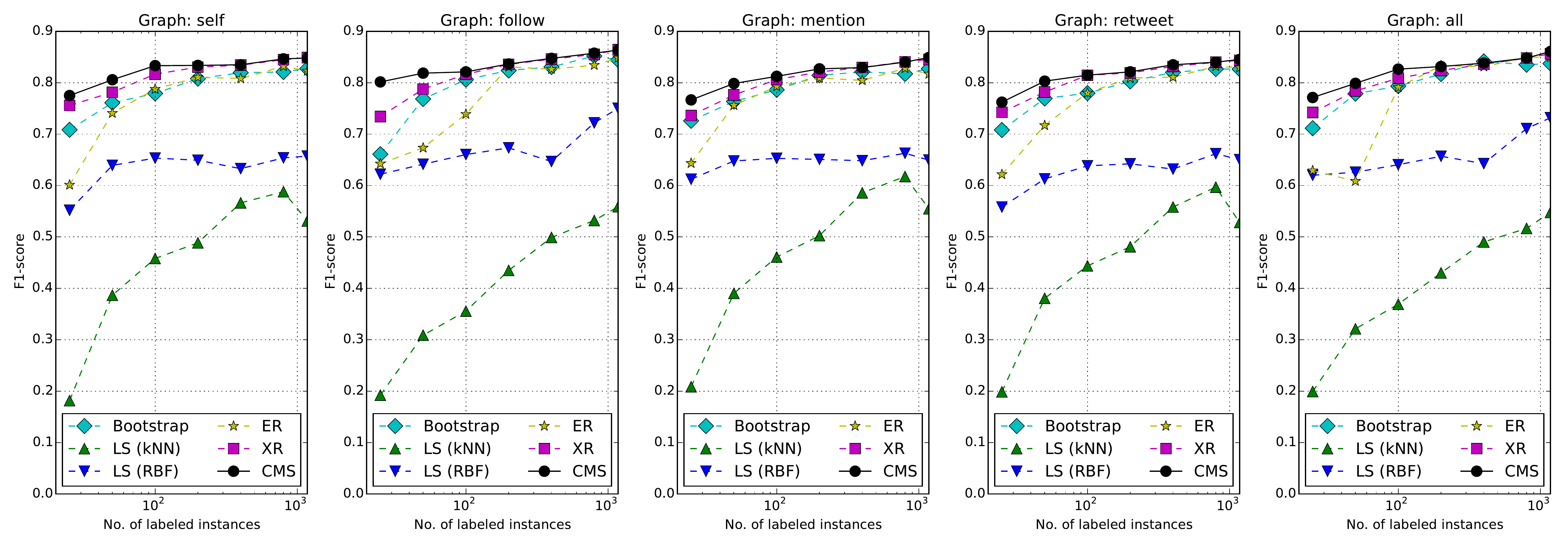}
\caption{Classification results for account type task with varying labeled data sizes and graph configurations}
\label{fig:account_sensitivity}
\end{figure*}

\begin{figure*}[!t]
\centering
\includegraphics[width=0.96\textwidth]{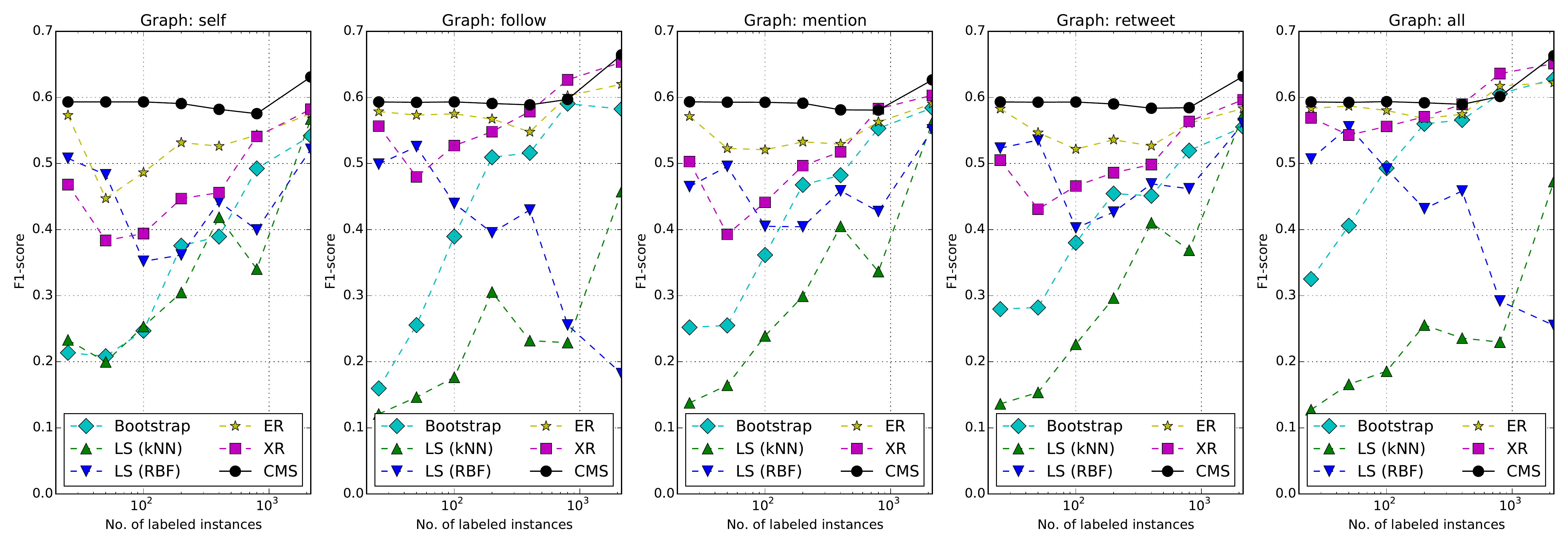}
\caption{Classification results for marital status task with varying labeled data sizes and graph configurations}
\label{fig:marital_sensitivity}
\end{figure*}

From Tables \ref{tab:account_benchmark} and \ref{tab:marital_benchmark}, we can see that our CSL method significantly outperforms the bootstrapping and LS methods, in terms of both $\text{\emph{F1-score}}$ and training time. This is true for all graph configurations. The $\text{\emph{F1-score}}$ of our method is also substantially better than that of the random guess baseline. It can be seen here that the ``high-confidence prediction is correct'' assumption of the bootstrapping method leads to a suboptimal performance (especially for the marital status task). It is also evident that the affinity assumption of the LS methods is inappropriate for our profiling tasks. In sum, these show that incorrect assumption about the data/task at hand in SSL can lead to mistakes that reinforce themselves.

Additionally, the results show that in general our CSL approach compares favourably to the ER and XR methods in terms of $\text{\emph{F1-score}}$, although there are cases where the performance difference is marginal (i.e., $p$-value $\geq 0.05$). Regardless, we will show later in a further sensitivity study (to be presented in Section \ref{sec:sensitivity}) that CSL is significantly more robust than the two methods. As for the training time, Tables \ref{tab:account_benchmark} and \ref{tab:marital_benchmark} show that CSL is as efficient as the ER and XR methods, but is an order of magnitude faster than the bootstrapping and LS methods. Finally, comparing the different graph configurations, we find that incorporating relational features from the follow graph alone already leads to $\text{\emph{F1-scores}}$ comparable to those using all the three graphs. This implies that the retweet and mention features are not as useful as the follow features. Nevertheless, unlike the LS methods (especially for the marital status task), using all three graphs in CSL does not significantly degrade the $\text{\emph{F1-scores}}$ (compared to using the follow graph alone). This suggests that CSL can make better use of MRF, even when noisy or less relevant relational features are used.

\subsection{Contribution of MRF and Unlabeled Data (RQ2)}
\label{sec:sensitivity}

To see the contributions of the MRF and CD formulations, further sensitivity studies were carried out by varying the graph configurations and number of labeled instances in the training data, respectively. We chose the number of labeled instances $L$ from $\{25, 50, 100, 200, 400, 800, \text{``full''}\}$, where ``full'' refers to $90\%$ of all the labeled instances, as already explained in Section \ref{sec:performance}. Fig. \ref{fig:account_sensitivity} and \ref{fig:marital_sensitivity} consolidate the results of our studies for account type and marital status tasks respectively. Note that the results at the right-hand extremes of the figures correspond those in Tables \ref{tab:account_benchmark} and \ref{tab:marital_benchmark}.

\begin{figure*}[!t]
\centering
\includegraphics[width=0.96\textwidth]{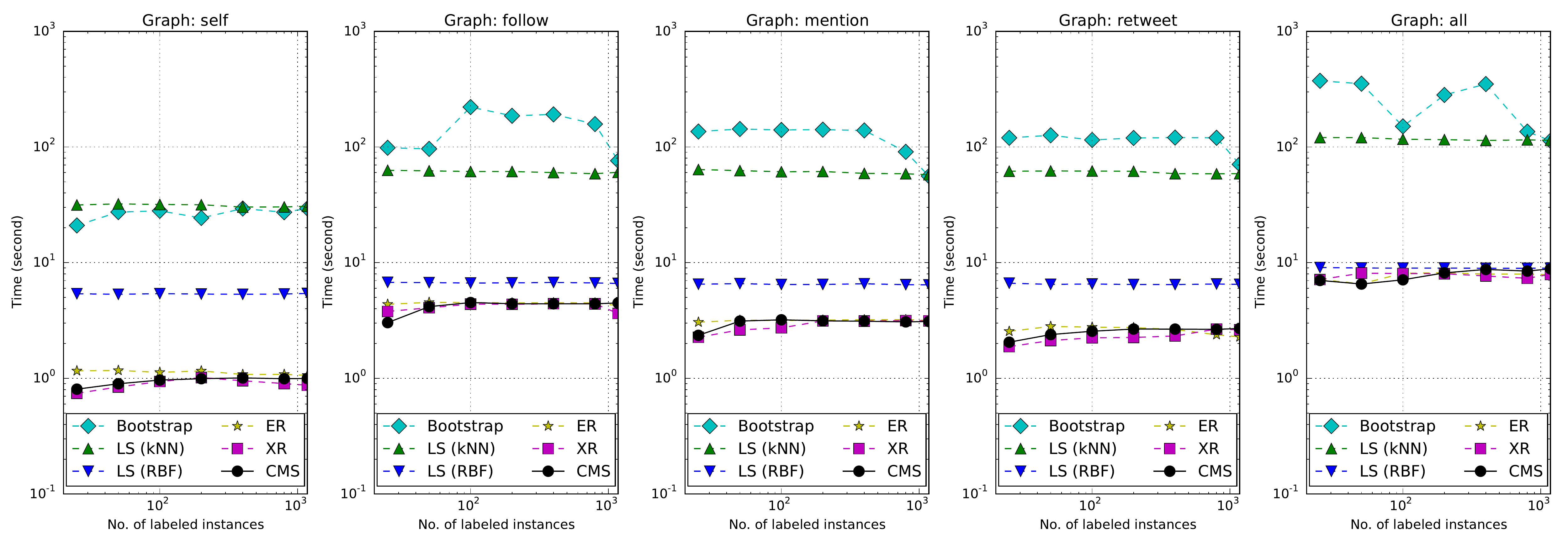}
\caption{Training time for account type with varying labeled data sizes and graph configurations}
\label{fig:account_time}
\end{figure*}

\begin{figure*}[!t]
\centering
\includegraphics[width=0.96\textwidth]{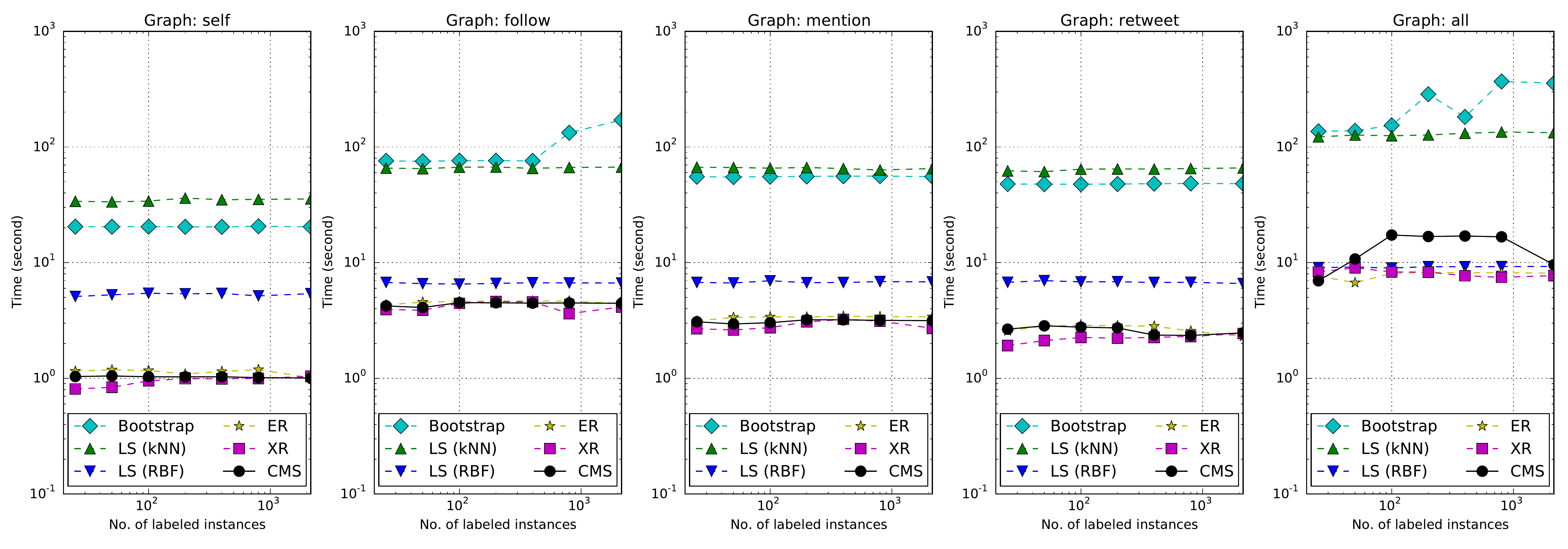}
\caption{Training time for marital status with varying labeled data sizes and graph configurations}
\label{fig:marital_time}
\end{figure*}

We can see here that the CD regularization in CSL yields more robust and consistent $\text{\emph{F1-scores}}$ than the ER and XR regularization, especially for small labeled data size. For example, even when $L = 25$, CSL is able to achieve $\text{\emph{F1-score}}$ not far from that obtained with $L = \text{``full''}$, for both profiling tasks. This is also verified by our Wilcoxon tests, all of which yielded $p$-values less than $0.05$. Moreover, our CSL method performs more consistently than the other methods under different graph configurations. As before, the retweet and mention features contribute less compared to the follow features. Finally, Fig. \ref{fig:account_time} and \ref{fig:marital_time} present the breakdown of training times of different methods. It is again shown that CSL is on par with ER and XR, but substantially faster than the bootstrapping and LS methods. All in all, these show that CSL exhibits both robustness and efficiency, making it more preferable than other methods for profiling tasks.

\subsection{Feature Importance Analysis (RQ3)}
\label{sec:importance}

\begin{figure*}[!t]
\centering
\includegraphics[width=0.88\textwidth]{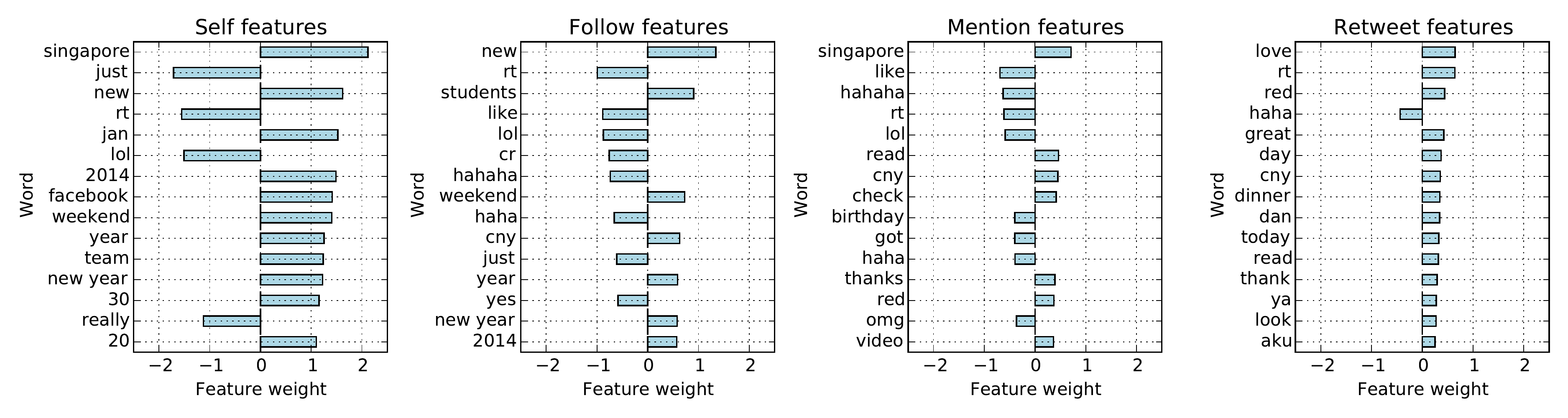}
\caption{Feature importance for account type classification}
\label{fig:account_importance}
\end{figure*}

\begin{figure*}[!t]
\centering
\includegraphics[width=0.88\textwidth]{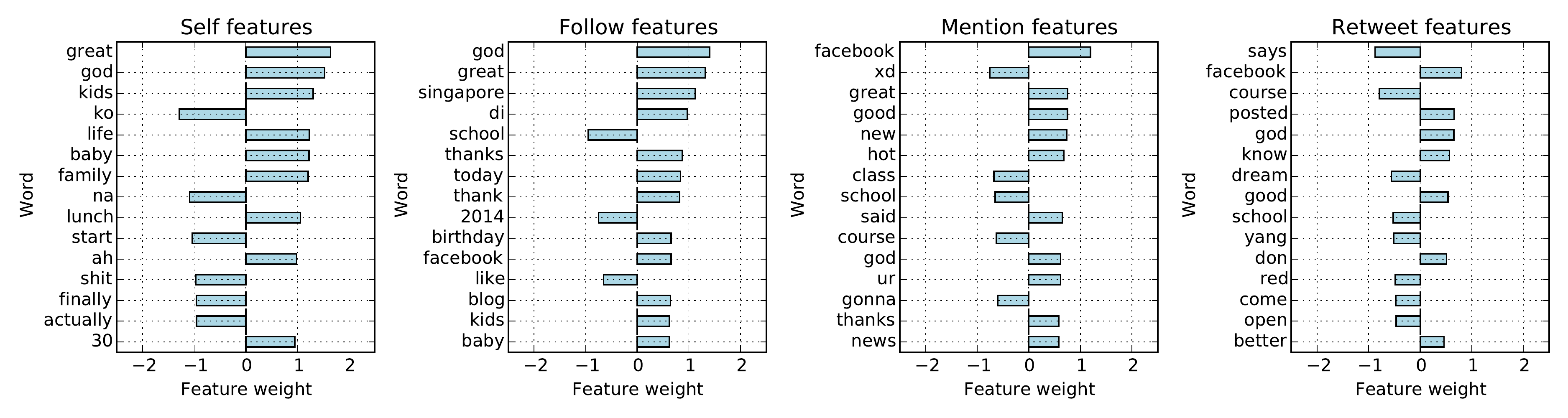}
\caption{Feature importance for marital status classification}
\label{fig:marital_importance}
\end{figure*}

We can now probe into the parameters of the trained CSL model, and investigate in details which features are the most discriminative for our user profiling tasks. Specifically, we assess the importance of the individual features by looking at the learned self weights $\theta_{j}$ and relational weights $\alpha_{m,j}$. 
Fig. \ref{fig:account_importance} and \ref{fig:marital_importance} present the top 15 features (i.e., having the largest \emph{absolute values} of $\theta_{j}$ and $\alpha_{m,j}$) for the account type and marital status profiling tasks respectively. The leftmost bar chart in each figure shows the self weights $\theta_{j}$, while the remaining charts show the relational weights $\alpha_{m,j}$ for the follow, mention, and retweet graphs respectively. We also note that a positive weight suggests that the respective feature is correlated with the positive label (i.e., ``organization'' or ``married''), whereas a negative weight corresponds to the negative label (i.e., ``personal'' or ``single'').

The results reveal several interesting insights that conform with our intuition. For example, in the account type profiling task, we can see that informal expressions such as ``just'', ``lol'' or ``haha'' are correlated with personal accounts, whereas organizations tend to be associated with more neutral words such as ``singapore'', ``new'', or ``cny'' (acronym for ``Chinese New Year''). Similarly for the marital status task, family-related words such as ``god'', ``great'' or ``kids'' are often correlated with married people, whereas single people consist of young adults and students, who like to use words such as ``school'', ``shit'', etc. In general, we also find that the feature correlation is consistent across different feature groups (e.g., when ``singapore'' in self features correlates with the organization account, the correlation also holds for ``singapore'' in mention features). 

Furthermore, it is evident from the overall weight magnitudes in Fig. \ref{fig:account_importance} and \ref{fig:marital_importance} that the self features are the most discriminative, though the other, relational features are still collectively useful. Among the three groups of relational features, we find that those derived from the follow graph are the most relevant. This can be attributed to the fact that the follow graph is more dense than the retweet or mention graph (see Table \ref{tab:count_dist}), thus yielding more informative features.

\begin{figure}[!t]
\centering
\includegraphics[width=0.51\textwidth]{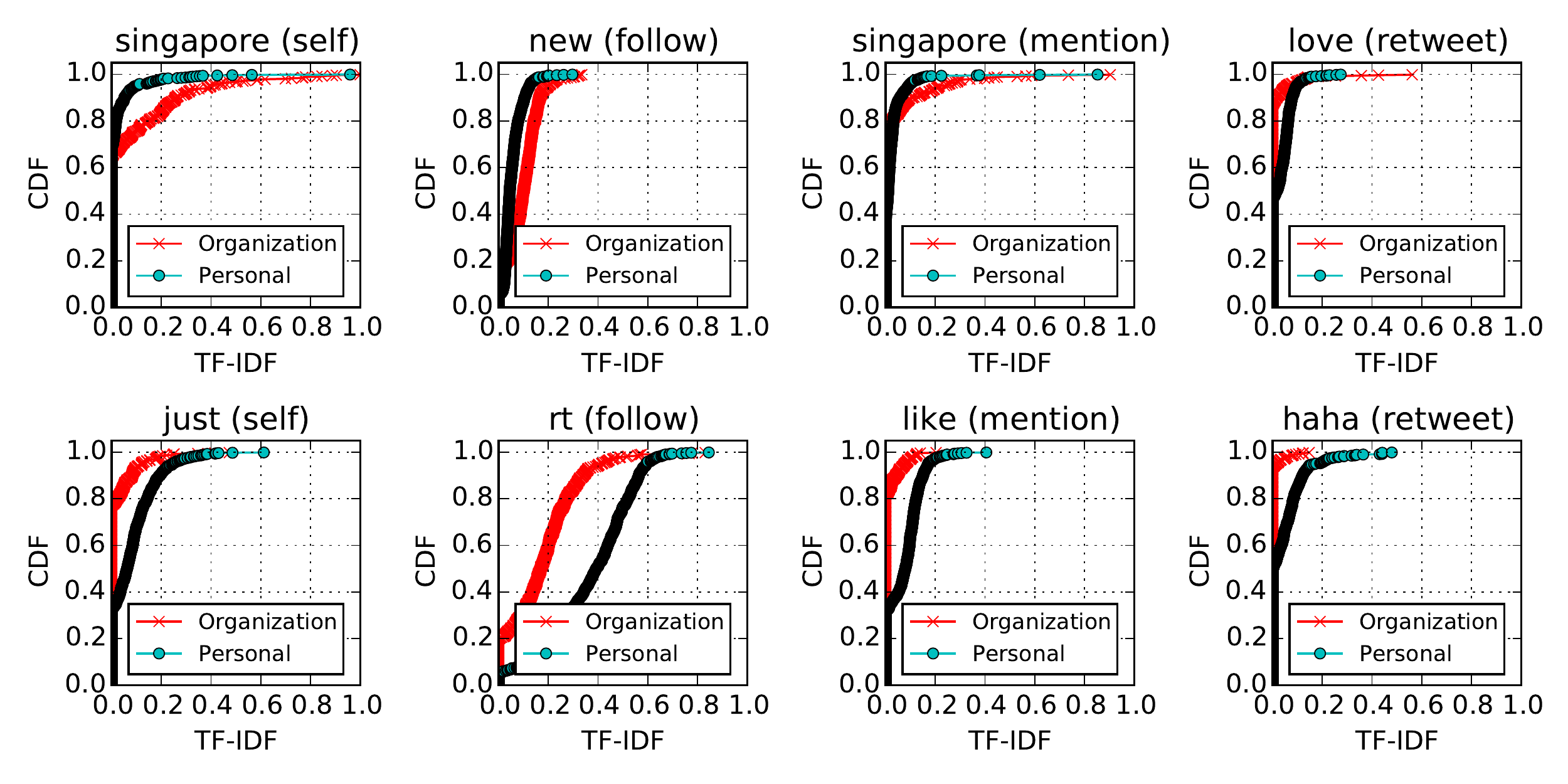}
\caption{Cumulative distribution of the top features for account type task}
\label{fig:account_hist}
\end{figure}

To examine the validity of the learned top features, we grouped all (labeled) instances according to the labels and examined the feature distributions for each label. Fig. \ref{fig:account_hist} shows the cumulative distribution function (CDF) of the top positive and negative features for each feature group in the account type task. An early increase in the CDF value implies a more skewed feature distribution, according to which we can then judge whether the learned positive or negative correlation is valid. For instance, the CDF of the word ``singapore'' is less skewed for organization accounts, suggesting that organizations use the word more often than personal accounts. Comparing Fig. \ref{fig:account_importance} with Fig. \ref{fig:account_hist}, we can conclude that, overall, the feature distributions conform with the learned feature correlations. The conclusion also holds for all the remaining features as well as the features for the marital status task. Owing to space constraints, however, we are not able to show the exhaustive results here.

\subsection{Generalization to Unseen Data (RQ4)}
\label{sec:generalization}

Further studies have been conducted to evaluate the ability of our learned model to generalize to novel, unlabeled data. To this end, we used our trained CSL model to predict for all unlabeled data, and picked the top $K$ positive instances with the highest prediction scores $\sigma_i$, as well as the top $K$ negative instances with the highest $(1 - \sigma_i)$. We then manually inspected all these instances to see how well the CSL predictions match with human judgments. To assess the model robustness, we varied $K$ from $20$ to $100$. For each class label, we recorded the number of correctly predicted instances ($TP$), that of incorrectly predicted instances ($FP$), and that of unclassifiable instances ($UC$). The UC refers to the case whereby we are unable to manually determine the actual label of an instance (i.e., a ``don't know'' answer). We then computed the precision at top $K$ ($Prec$) by excluding the unclassifiable instances, i.e., $Prec = \frac{TP}{K - UC}$.

Tables \ref{tab:account_confusion} and \ref{tab:marital_confusion} summarize the results for account type and  marital status tasks respectively. Overall, we can see that the learned CSL model can predict for both the account type and marital status pretty well, which is evident from the fairly low $FP$ and high $Prec$ results. Meanwhile, the $UC$ numbers are generally low, except for the ``married'' label of the marital status task. The latter suggests that determining whether a person is married is a difficult task, even for humans. Nonetheless, the overall results demonstrate that our method has good generalization abilities.

\subsection{Inference on Larger User Population (RQ5)}
\label{sec:qualitative}

The final part of our empirical studies consists of making inference on a larger population of SGTwitter users. To this end, we carried out quantitative and qualitative analyses on the predictions made for all unlabeled data. Our quantitative study involves comparing the label distributions in the labeled dataset (as per Table \ref{tab:label_dist}) with the predicted distributions as inferred by CSL on the unlabeled dataset. Table \ref{tab:label_comparison} shows the results. We find that the predicted distributions are much more imbalanced than the distributions of the labeled data. Although we cannot fully verify this observation (due to the need to label all  100K samples), it is reasonable to expect that the larger SGTwitter population would consist of more personal accounts than organization ones, and more single users than married ones. The distribution difference also suggests that the larger population contains new cases that are not previously captured in the labeled data, and our method can generalize to these cases fairly well. 

\begin{table}[!t]
\scriptsize
\centering
\caption{Top $K$ predictions on unseen data instances for account type task}
\begin{tabular}{l|l|c|c|c|c|c}
\hline
Label	& Metric			& Top 20	& Top 40	& Top 60	& Top 80	& Top 100\\
\hline
Organization	& $TP$	& 18	& 36	& 56	& 72	& 89\\
				& $FP$	& 2		& 4		& 4		& 8		& 11\\
				& $UC$    & 0		& 0		& 0		& 0		& 0\\
				& $Prec$	& 0.9000	& 0.9000	& 0.9333 	& 0.9000	& 0.8900\\
\hline
Personal	    & $TP$	& 18	& 37	& 55	& 73	& 90\\
				& $FP$	& 0		& 0		& 0		& 0		& 0\\
				& $UC$	& 2		& 3		& 5		& 7		& 10\\
				& $Prec$	& 1.0000	& 1.0000	& 1.0000	& 1.0000	& 1.0000\\
\hline
\multicolumn{7}{l}{$TP$: true positives, $FP$: false positives, $UC$: unclassifiable, $Prec$: precision}
\end{tabular}
\label{tab:account_confusion}
\end{table}

\begin{table}[!t]
\scriptsize
\centering
\caption{Top $K$ predictions on unseen data instances for marital status task}
\begin{tabular}{l|l|c|c|c|c|c}
\hline
Label	& Metric	& Top 20	& Top 40	& Top 60	& Top 80	& Top 100\\
\hline
Married	& TP		& 9		& 19	& 24	& 33	& 39\\
		& FP 		& 0		& 0		& 2		& 4		& 6\\
		& UC		& 11	& 21	& 33	& 42	& 54\\
		& Prec		& 1.0000	& 1.0000	& 0.9231	& 0.8919	& 0.8667\\
\hline
Single	& TP		& 16	& 34	& 53	& 69	& 88\\
		& FP 		& 0		& 0		& 0		& 0		& 0\\
		& UC		& 4		& 6		& 7		& 11	& 12\\
		& Prec		& 1.0000	& 1.0000	& 1.0000	& 1.0000	& 1.0000\\
\hline
\multicolumn{7}{l}{$TP$: true positives, $FP$: false positives, $UC$: unclassifiable, $Prec$: precision}
\end{tabular}
\label{tab:marital_confusion}
\end{table}

\begin{table}[!t]
\scriptsize
\centering
\caption{Comparison of label distributions}
\begin{tabular}{l|l|c|l|c}
\hline
Dataset         & \multicolumn{2}{c|}{Account type} & \multicolumn{2}{c}{Marital status}\\
\cline{2-5}
     		 & Label        & \#users 		  & Label     & \#users\\
\hline
Labeled      & Personal 	& 794 (60.7\%)    & Single 	  & 1,304 (56.4\%)\\
			 & Organization & 514 (39.3\%)    & Married   & 1,009 (43.6\%)\\
\hline
Unlabeled    & Personal     & 88,101 (88.0\%) & Single    & 62,596 (62.4\%) \\
			 & Organization & 12,396 (12.0\%) & Maried    & 37,901 (37.6\%) \\
\hline
\end{tabular}
\label{tab:label_comparison}
\end{table}

\begin{figure}[!t]
\centering
\includegraphics[width=0.45\textwidth]{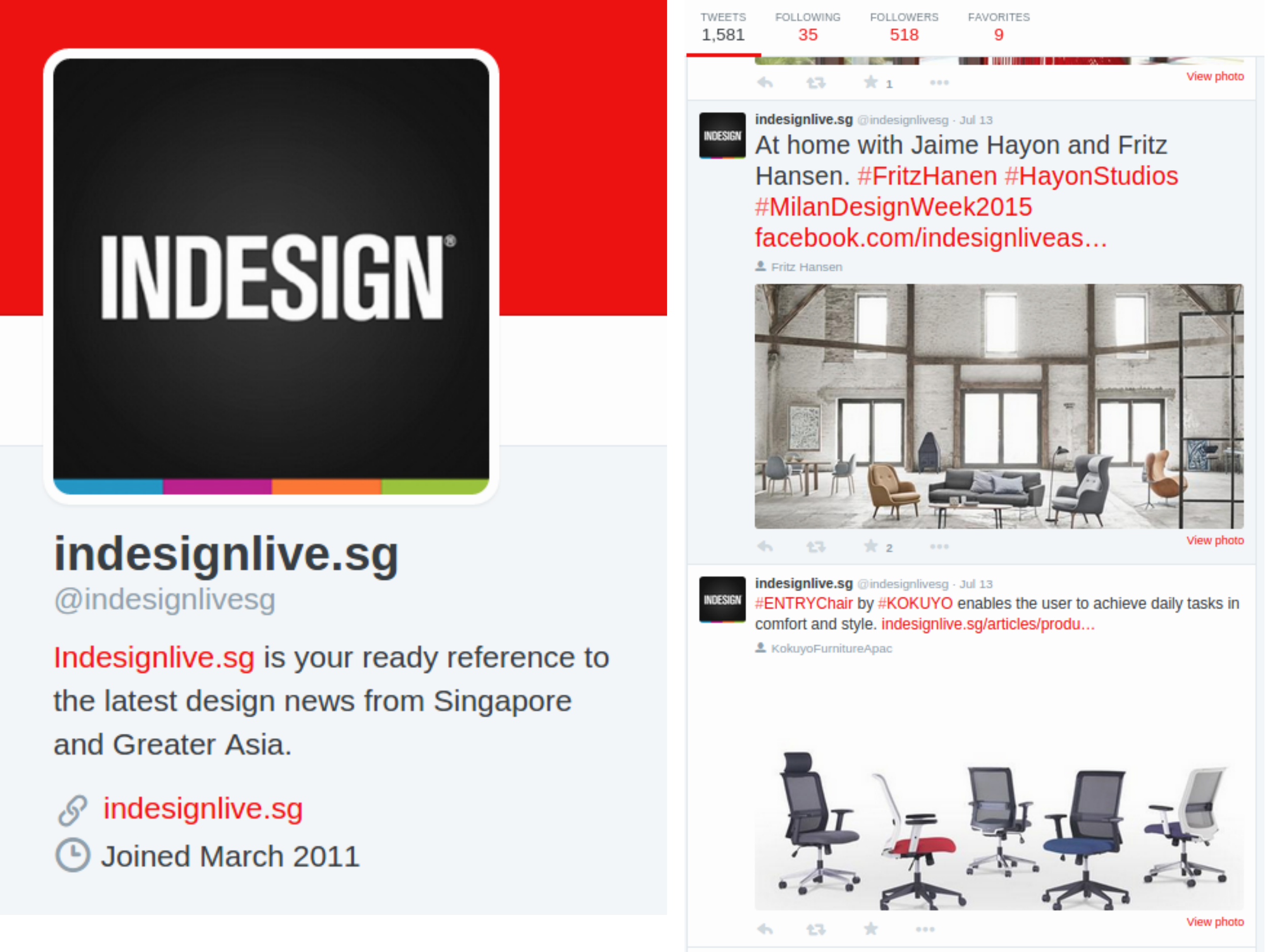}
\caption{Example of top organization account}
\label{fig:qualitative}
\end{figure}

Further qualitative analysis on the individual users reveals additional insights about the (larger) SGTwitter population. Fig. \ref{fig:qualitative} shows the screenshot of a top-predicted organization account that is not previously captured in the labeled dataset. In particular, the profile description of the account has an URL with a new suffix ``.sg'', which is not part of the suffices used to derive the labeled dataset (i.e., ``.com.sg'', ``.edu.sg'' and ``.gov.sg''; see Section \ref{sec:attribute}). This shows that our CSL method is able to properly predict for novel instances, based on content (word) features alone. Similarly for the marital status task, we found several interesting insights (not shown here due to space limitation). For instance, some of the top-predicted married users never use the keywords/phrases listed for the labeled data (see again Section \ref{sec:attribute}) in their profile descriptions, but their profile pictures clearly show that they have a spouse or children. 

%% file: conclusion.tex
\section{Conclusion}
\label{sec:conclusion}

In this paper, we put forward a novel CSL approach for modeling/profiling the attributes of users in social media. The proposed approach provides a principled and efficient solution to the novel problem of simultaneously exploiting multiple types of social relationship and large pool of unlabeled data in user profiling tasks. The centerpiece of the proposed CSL approach is to first expand the input space by generically constructing a set of MRF features that capture different types of relationship, and then perform the CD regularization to establish convex semi-supervised learning using unlabeled data. The experimental results on Singapore Twitter users have demonstrated the accuracy, robustness, efficiency, as well as interpretability traits of our approach in profiling the user attributes.

Moving forward, we wish to extend our methodology to more challenging profiling tasks involving multiple social networks (e.g., Facebook, Foursquare, etc.). We also plan to develop a multi-task learning framework that can infer multiple user attributes jointly by modeling their dependencies (e.g., correlation between age group and marital status).